\documentclass[aps,prd, onecolumn,nofootinbib,showpacs,showkeys,12pt]{revtex4-1}

\usepackage{graphicx,color}
\usepackage{enumerate}
\usepackage{amsmath,amssymb,amsthm}
\usepackage[paperwidth=210mm,paperheight=297mm,centering,hmargin=20mm,vmargin=20mm]{geometry}
\usepackage{float}
\usepackage{mathrsfs}
\usepackage{hyperref}
\usepackage{physics}
\usepackage[T1]{fontenc}
\usepackage{slashed}
\usepackage[usenames,dvipsnames]{xcolor}
    \definecolor{officegreen}{rgb}{0.0, 0.5, 0.0}

\usepackage{accents}
\newcommand*{\dt}[1]{%
  \accentset{\mbox{\large\bfseries .}}{#1}}

\renewcommand{\i}{\text{i}}
\renewcommand{\d}{\text{d}}
\newcommand{\e}{\text{e}}

\newcommand{\be}{\begin{equation}}
\newcommand{\ee}{\end{equation}}
\newcommand{\beq}{\begin{eqnarray}}
\newcommand{\eeq}{\end{eqnarray}}

\date{\today}

\begin{document}

\title{Observational Imprints of Our Lost Twin Anti-Universe}

\author{Samuel Barroso Bellido}
\email{Main author}
\email{samuel.barroso-bellido@usz.edu.pl}
\affiliation{Institute of Physics, University of Szczecin, Wielkopolska 15, 70-451 Szczecin, Poland}

\author{Mariusz P. D\c{a}browski}
\email{Mariusz.Dabrowski@usz.edu.pl}
\affiliation{Institute of Physics, University of Szczecin, Wielkopolska 15, 70-451 Szczecin, Poland}
\affiliation{National Centre for Nuclear Research, Andrzeja So{\l}tana 7, 05-400 Otwock, Poland}
\affiliation{Copernicus Center for Interdisciplinary Studies, Szczepa\'nska 1/5, 31-011 Krak\'ow, Poland}

\begin{abstract}
%We present the results of the effects on the Cosmic Microwave Background of the interuniversal interaction given by the entanglement between pairs of universes in the multiverse. It is based on the Third Quantization picture of Canonical Quantum Gravity and the recent discovery that the entanglement entropy is related to the Hubble parameter such that it diverges at the critical points of the classical evolution of the universes.
We consider observational consequences of the entanglement between our universe and a hypothetical twin anti-universe in the third quantization scheme of the canonical quantum gravity. Based on our previous investigations we select some special form of the interuniversal interaction which allows the entanglement entropy of the pair of universes to diverge at some critical points of their classical evolution. We find that the modification of the cosmic microwave background (CMB) power spectrum due to the entanglement with our twin anti-universe is enlarged for small modes $k$ and small multipole numbers $l$ with the Planck constraint onto the interaction coupling constant  $\lambda_o\lesssim\mathcal{O}(10^{-56})$. Some other coupling functions which allow more critical points are also briefly commented on in the context of their observational effect on CMB and other observations. 
\end{abstract}

\maketitle

\section{Introduction}

Year by year, the multiverse scenario reaches more and more important status in theoretical physics \cite{Multi1,Multi2,Multi3,universe2019,Multi4,Multi5,Multi6,Multi7,Multi8,BalcerzakBell}. The question whether the multiverse exists or not is still open and under debate. In Ref. \cite{Interacting}, a level III multiverse scenario \cite{Tegmark} is put to the test for which the imprints on the Cosmic Microwave Background (CMB) spectrum are found. It seems then appealing to apply its strategy as viable and proceed to find the observable signals from the CMB for other scenarios.  

The method relies on the third quantization scheme \cite{3Q1,3Q2,Giddings1988,3QSalva} of canonical quantum gravity \cite{DeWitt1,KieferBook} which now is widely considered the field theory of universes over the multiverse, that, assuming they are Friedmann-Lemaitre-Robertson-Walker (FLRW) universes, each universe is described by its wave function $\Psi(a,\{\phi\})$, where $a$ is the scale factor and $\{\phi\}$ the matter content. Such wave function fulfils its own renowned Wheeler-DeWitt (WDW) equation $\mathcal{H}\Psi=0$. In that kind of the multiverse, the individual universes interact as particles of any other field theory. In particular, they follow the most natural process of quantum field theory which is the pair creation of universes \cite{Babies1,Babies2} ''from nothing'' as an analogy with particle-antiparticle pair creation \cite{SchwingerEffect}. Unlike in Ref. \cite{Interacting}, where the number of interacting universes is arbitrary and large, we will just consider entangled and interacting pair of universes with no other interactions or environment. 

Besides, a detail arose when the entanglement entropy of such pair of universes was calculated \cite{Cyclic,Critical,SBB2}. It was found to be divergent close to any critical point along the classical evolution of the universes, or equivalently, when the Hubble parameter $H$ vanishes (turning points of possible cyclic evolution - cf. Ref. \cite{kiefer95,MPD95}). As we know from quantum mechanics, the Hamiltonian description of such a bipartite system with a certain interaction between them can be written as
\begin{equation}\label{HamSum}
    \mathcal{H}=\mathcal{H}_1+\mathcal{H}_2+\mathcal{H}_{\text{int}},
\end{equation}
where $\mathcal{H}_i$ is the free Hamiltonian of each subsystem, and $\mathcal{H}_{\text{int}}$ is the contribution to the Hamiltonian by their interaction. Since the entanglement entropy diverges when the Hubble parameter vanishes, the interaction must also diverge. One is then compelled to write the Hamiltonian of interaction as the function of the Hubble parameter and not solely as the function of the scale factor only $a$, as it was considered in Ref. \cite{Interacting}.

In order to find the Hamiltonian of the third quantized theory, we need to find the action for which each universe fulfils its WDW equation and interacts with the other. In order to do so, we find the individual actions, sum them up, and include the yet unknown interaction term. Our proposal is to include rather general form for the interaction Hamiltonian based on the nearest neighbour interaction process
\begin{equation}\label{Hint}
    \mathcal{H}_{\text{int}}(a,H)=-a\lambda^2(a,H)(\Psi_2-\Psi_1)^{\ast}(\Psi_2-\Psi_1),
\end{equation}
where $\lambda^2(a,H)$ is a coupling function whose constraints are to be based on two conditions. Firstly, this function must depend on the Hubble parameter in such a way that the coupling diverges when the Hubble parameter vanishes, i.e. 
\begin{equation}
\label{Limit}
    \lim_{H\to0}\lambda(a,H)\to\infty,
\end{equation}
and secondly (cf. Ref. \cite{Vacuum}), the probability of the false vacuum to decay which is computed for such kind of interaction
\begin{equation}
\label{Cond}
\mathcal{P}\sim\exp\left\{-\left[\frac{a^4}{\lambda^2(a,H)}\right]^3\right\},
\end{equation}
is suppressed at large values of the scale factor $a \to \infty$.

The development to find the Wheeler-DeWitt equation and the semiclassical Friedmann equation which corresponds to that model is the purpose of Section \ref{S1}. Using the new dynamics, we find the spectra of the CMB in Section \ref{S3}. A comment of the choice of a certain coupling function $\lambda^2(\alpha,\phi)$ is given in Section \ref{SDE}. The conclusion follows in Section \ref{SConc}.

\section{Third Quantized Wheeler-DeWitt Equation and the Semiclassical Friedmann Equation}
\label{S1}

Considering a pair of flat universes containing a single homogeneous scalar field $\phi$, the mode expansion of its perturbations $v_k$, and the perturbations of the metric, the Lagrangian of each universe of the pair can be expressed, in terms of the conformal time $\d\eta=\d t/a(t)$, like \cite{Interacting,Morais}
\begin{equation}\label{Lag1}
    \mathcal{L}=\frac{1}{2}\left[-(a')^2+a^2(\phi')^2-2a^4V(\phi)+\sum_k(v_k'{v_{k}^{\ast}}'+\omega^2_kv_kv^{\ast}_k)\right],
\end{equation}
where the prime denotes the derivative with respect to $\eta$, we have assumed $3/4\pi G=1$, $G$ is the gravitational constant, $V(\phi)$ is the potential of the scalar field, and the frequency is given by
\begin{equation}
    \omega^2_k=k^2-\frac{z''}{z}, \qquad\qquad z=a\frac{\phi'}{\mathfrak{H}}, \qquad \qquad \mathfrak{H}=\frac{a'}{a},
\end{equation} 
where $z$ is called the Mukhanov-Sasaki variable, and the perturbations $v_k$ are the solutions to the Mukhanov-Sasaki equation
\begin{equation}\label{MSEq}
    v''_k+\left(k^2-\frac{z''}{z}\right)v_k=0.
\end{equation}

The conjugated momenta are found from Eq. (\ref{Lag1}) to be
\begin{equation}
    p_a=-a',\qquad \qquad p_{\phi}=a^2\phi',\qquad \qquad p_{v_k}={v_k^{\ast}}',
\end{equation}
and a Legendre transformation yields the Hamiltonian
\begin{equation}\label{Ham1}
    \mathcal{H}=\frac12\left[-p_a^2+2a^4\rho(\phi)+\sum_k\left(p_{v_k}p_{v^{\ast}_k}-\omega^2_kv_kv_k^{\ast}\right)\right],
\end{equation}
where we kept the scalar field as classical through the relation \cite{Weinberg}
\begin{equation}
    \rho(\phi)=\frac12\frac{(\phi')^2}{a^2}+V(\phi).
\end{equation}

In order to find the Wheeler-DeWitt equation $\mathcal{H}\Psi=0$ of a single universe, where $\Psi(a,\{v_k\})$ is the wave function of each one, we now quantize the momenta, for a certain factor ordering, as
\begin{equation}
    p_a^2=-\frac{1}{a}\frac{\partial}{\partial a}\left(a\frac{\partial}{\partial a}\right),\qquad \qquad p_{v_k}=-\i\frac{\partial}{\partial v_k},
\end{equation}
and replace them into the Hamiltonian (\ref{Ham1}), so that the kinetic part could be expressed covariantly as the Laplace-Beltrami operator. Thus, one finds the Wheeler-DeWitt equation
\begin{equation}\label{WDW1}
    \left[\frac{1}{a}\frac{\partial}{\partial a}\left(a\frac{\partial}{\partial a}\right)
    +2a^4\rho(\phi)+\sum_k\left(-\frac{\partial^2}{\partial v_k^2}+\omega^2_kv_k^2\right)\right]\Psi(a,\{v_k\})=0,
\end{equation}
where $\Psi(a,\{v_k\})$ is the wave function of the universe, which is a function on the scale factor and the modes of the perturbations.

In the third quantization picture, Eq. (\ref{WDW1}) is the equation of motion for a certain field $\Psi^{\ast}$, meaning that the universe described by $\Psi$ in Eq. (\ref{WDW1}) is just an excitation of a field of universes. The action from where we recover it, reads
\begin{equation}
    S_{3Q}=\int\d a\prod_k\d v_k a\left[-\frac{\partial\Psi^{\ast}}{\partial a}\frac{\partial\Psi}{\partial a}+2a^4\rho(\phi)\Psi^{\ast}\Psi
    +\sum_k\left(\frac{\partial\Psi^{\ast}}{\partial v_k}\frac{\partial\Psi}{\partial v_k}+\omega^2_k v_k^2\Psi^{\ast}\Psi\right)\right].
\end{equation}
Identifying the momentum of the universe as
\begin{equation}
    P_{\Psi}=-a\frac{\partial\Psi}{\partial a},
\end{equation}
we find the third quantized Hamiltonian 
\begin{equation}
    \mathcal{H}_{3Q}
    =a\left[-\frac{1}{a^2}P_{\Psi}P_{\Psi^{\ast}}
    -2a^4\rho(\phi)\Psi^{\ast}\Psi-\sum_k\left(\frac{\partial\Psi^{\ast}}{\partial v_k}\frac{\partial\Psi}{\partial v_k}+\omega^2_k v_k^2\Psi^{\ast}\Psi\right)\right],
\end{equation}
which accounts for one of the universes of the bipartite system of two interacting universes we want to focus on. The whole system is described by a Hamiltonian like the one in Eq. (\ref{HamSum})
\begin{equation}\label{Hmulti}
\mathcal{H}_{\text{pair}}=\mathcal{H}_{3Q}^1+\mathcal{H}_{3Q}^2-a\lambda^2(a,H)(\Psi_2-\Psi_1)^{\ast}(\Psi_2-\Psi_1),    
\end{equation}
where the interaction term has been taken as in Eq. (\ref{Hint}), controlled by a coupling function $\lambda(a,H)$ which here is the clue to justify the divergence of the entanglement entropy at the critical points of the classical evolution as it was recently found in Ref. \cite{Critical}, and to maintain the vacuum stability.

To simplify a bit the calculations, we expand the wave functions in its frequency modes like
\begin{equation}\label{Fourier}
    \Psi_{1,2}=\frac{1}{\sqrt{2}}
    \left(\widetilde{\Psi}_2\mp\widetilde{\Psi}_1\right),
    \qquad \qquad 
     P_{\Psi_{1,2}}=\frac{1}{\sqrt{2}}
     \left(\widetilde{P}_{\widetilde{\Psi}_2}\mp\widetilde{P}_{\widetilde{\Psi}_1}\right),
\end{equation}
since there are only two universes. Substituting (\ref{Fourier}) into (\ref{Hmulti}), we get the third quantized Hamiltonian of the system, which is
\begin{equation}
    \widetilde{\mathcal{H}}_{3Q}^{\text{pair}}
    =\widetilde{\mathcal{H}}_{3Q}^1
    +\widetilde{\mathcal{H}}_{3Q}^2-2a\lambda^2(a,H)\widetilde{\Psi}^{\ast}_1\widetilde{\Psi}_1,
\end{equation}
where
\begin{equation}
    \widetilde{\mathcal{H}}_{3Q}^l
    =a\left[-\frac{1}{a^2}
    \widetilde{P}_{\widetilde{\Psi}_l}\widetilde{P}^{\ast}_{\widetilde{\Psi}_l}
    -2a^4\rho(\phi)\widetilde{\Psi}_l^{\ast}\widetilde{\Psi}_l-\sum_k\left(\frac{\partial \widetilde{\Psi}_l^{\ast}}{\partial v_k}\frac{\partial \widetilde{\Psi}_l}{\partial v_k}+\omega^2_k v_k^2\widetilde{\Psi}_l^{\ast}\widetilde{\Psi}_l\right)\right], \qquad l=\{1,2\}.
    \end{equation}

The equation of motion for $\widetilde{\Psi}_l^{\ast}$, which is the Wheeler-DeWitt equation for $\widetilde{\Psi}_l$, after a parameterization of the scale factor as $\alpha=\ln(a)$, is
\begin{equation}\label{WDWFinal}
    \left[\e^{-2\alpha}\frac{\partial^2}{\partial\alpha^2} +2\e^{4\alpha}\rho(\phi)
    -\frac{\partial^2}{\partial v_k^2}
    +\omega^2_k v_k^2+2\lambda^2(a,H)\delta(l-1)\right]\widetilde{\Psi}_l(\alpha,\{v_k\})=0,
\end{equation}
where $\delta(l-1)$ is a Dirac delta function. Here, $\widetilde{\Psi}_l$ is the wave function with mode $l$ of the bipartite system. From (\ref{WDWFinal}), we can see that the mode $l=2$ does not feel the interaction at all, hence the dynamics of the classical universe related to it is equivalent to a non-interacting one. The question is in what combination of them is our universe, and this is impossible to answer. We will, therefore, focus our attention over the mode $l=1$ for which the dynamics differs, and from now on, no index will be used.

Using the semiclassical ansatz
\begin{equation}
    \widetilde{\Psi}(\alpha,\{v_k\})=\e^{\i S_{0}(\alpha)}\prod_k\psi_{k}(\alpha,\{v_k\}),
\end{equation}
we find the approximate relation\footnote{For the explicit calculation, see the  Appendix A of the analogous case in Ref. \cite{Interacting}.}
\begin{equation}\label{HJEq}
    -\e^{-2\alpha}\left(\frac{\partial S_0}{\partial\alpha}\right)^2
    =-2\e^{4\alpha}\rho(\phi)-2\lambda^2(a,H).
\end{equation}
Recognizing (\ref{HJEq}) it as the Hamilton-Jacobi equation for the slowly moving background, we use it together with 
\begin{equation}
    \frac{\partial S_0}{\partial a}=p_a=-a\frac{\d a}{\d t},
\end{equation}
in order to obtain the Friedmann equation
\begin{equation}\label{FE}
    H^2=2\rho(\phi)+2\frac{\lambda^2(a,H)}{a^4},
\end{equation}
where we have recovered the scale factor and the cosmological time $t$. For the fluctuations $v_k$, it is found that they obey the Schr\"odinger-like equation \cite{Interacting}, as expected.

\section{Observational Consequences: CMB Angular Power Spectrum}\label{S3}

The new dynamics ruled by the Friedmann equation (\ref{FE}) which takes into account a small interaction between our universe and its twin by means of the coupling function $\lambda(a,H)$, may have an impact on the anisotropies of the CMB. The intention of this section is to find out how it makes a difference from the best fit found by Planck \cite{PlanckSpectrum}, where $\Lambda$CDM model is assumed.  We will consider pretty general coupling function fulfilling the condition (\ref{Limit}), which reads
\begin{equation}\label{Coupling}
    \lambda^2(a,H)=\frac{\lambda_o}{2}\frac{a^q}{H^n} = \frac{\lambda_o}{2}\frac{a^{q+n}}{\dot{a}^n},
\end{equation}
where $n>0$, and $q$ are some real constants, and $\lambda_o$ is a constant small enough to see the interaction between the pair as a  perturbation for large values of the scale factor, whose units are $[T^{-2-n}]$. However, a priori it is not valid to perform any approximation around $\lambda_o$ since such a term contributes mainly to the dynamics close to the initial singularity.

In order to also adopt other types of extreme points of the universe discussed in Ref. \cite{Critical} such as exotic singularities \cite{Krakow2014,GRG2018} which may have regular energy density but irregular the pressure and its derivatives, one may also postulate some more general ansatz 
\begin{equation}\label{Couplingq}
    \lambda^2(a,H,\bar{q})=\frac{\lambda_o}{2}\frac{a^q}{H^n \bar{q}^m} = (-1)^m \frac{\lambda_o}{2}\frac{a^{q+n-m}}{\dot{a}^{n-2m} \ddot{a}^m},
\end{equation}
where $\bar{q} \equiv - (\ddot{a} a)/\dot{a}^2$ is the deceleration parameter, and the constant $m>0$. For example, in a sudden future singularity (SFS) \cite{BarrowSFS} the scale factor $a = a_s =$ const., and its first derivative (proportional to the energy density) $\dot{a} = \dot{a}_s=$ const., while the second derivative (proportional to the pressure) $\ddot{a} \to - \infty$. This procedure may further be extended into the singularities in higher derivatives of pressure (for example generalized sudden future singularities (GSFS) \cite{GSFS-Barrow}) by inserting higher-order kinematic quantities in the denominator such as jerk, snap, and pop \cite{Caldwell_2004,Dunajski_2008,MPD2005} which would involve higher order derivatives of the scale factor into (\ref{Couplingq}). 

However, just focusing our attention on the coupling function (\ref{Coupling}), in order to recover the spectrum, we should use a scalar field whose equation of state is $p=\omega\rho$, where $\omega$ is expected to be very close to $\omega_{\Lambda}=-1$, since the expansion must be almost de Sitter \cite{Bassett}. Let us then write
\begin{equation}
    \omega=-1+\frac{2\alpha}{3},\qquad \qquad \alpha\gtrsim0.
\end{equation}
For a universe with a scalar field fulfilling the above condition, the density \cite{Weinberg} scales like  $\rho(\phi)=(1/2)H_{\text{dS}}^2 (a_d/a)^{2\alpha}$, and thus Eq. (\ref{FE}) is written as
\begin{equation}\label{HLambda}
    H^2=H_{\text{dS}}^2\left[\left(\frac{a_d}{a}\right)^{2\alpha}
    +\frac{\lambda_o}{H_{\text{dS}}^2H^{n}a^{4-q}}\right],
\end{equation}
where $a_d$ is a constant to be determined.

The condition for the probability (\ref{Cond}) can be fulfilled, when $4-q-n>0$. Hence, at the very beginning, the universe is dominated by a term which is {\it essentially different} from the one of Ref. \cite{Interacting}, since it depends on the Hubble parameter in the denominator. 

The analysis of an arbitrary $q$ and $n$ is left for future research. Here we will only consider the values $q=1$ and $n=1$, which is the extreme case since the Hubble parameter goes like $1/H$ in the interaction term, and it is positive during inflation. Therefore, any other value of $n$ will decrease the interaction and the effects would be less noticeable. This case is one of the family of cases for which the analysis is simplified fulfilling the condition $2-q-n=0$, so that the Friedmann equation can be written as
\begin{equation}\label{X80}
    \dt{a}^{n+2}=H_{\text{dS}}^2\dt{a}^na_d^{2\alpha}a^{2(1-\alpha)}+\lambda_o.
\end{equation} 
The asymptotic behaviour one can obtain from Eq. (\ref{X80}) of the scale factor close to the singularity is 
\begin{equation}\label{InitialExp}
    a(t)\approx \sqrt[3]{\lambda_o}t,
\end{equation}
which is an extremely slow motion of the (Milne) universe which appears on the edge of inflationary condition $\ddot{a} \geq 0$. Using Eq. (\ref{InitialExp}) as feedback for Eq. (\ref{HLambda}), we infer that a more precise expansion is still slightly accelerated.

Even if there is a short period before inflation, it should have been in a constant thermal equilibrium and no much physics is expected to happen. Indeed, one finds that $(aH)\sim\sqrt[3]{\lambda_o}$ there, which is constant throughout the initial stage. Furthermore, the dynamics of the very beginning of the universe is not perfectly obtained from canonical quantum gravity due to the problem of the factor ordering \cite{KieferBook}. For these reasons, we will not repeat the method used in Ref. \cite{Interacting}, and we will neglect the effects at the very beginning of the universe assuming that $\lambda_o$ can be taken as a perturbation at any point of the evolution. 

The equation (\ref{HLambda}) can be expanded at first order around $\lambda_o$ as
\begin{equation}\label{Happ}
    H\approx
    H_{\text{dS}}\left(\frac{a_d}{a}\right)^{\alpha}
    +\frac{\lambda_o}{2H_{\text{dS}}Ha^{3-\alpha}a_d^{\alpha}}
    \approx
    H_{\text{dS}}\left(\frac{a_d}{a}\right)^{\alpha}
    +\frac{\lambda_o}{2H_{\text{dS}}^2a^{3}}.
\end{equation}
Trying the ansatz
\begin{equation}
    a(t)=a_o(t)\left[1+\xi(t,\lambda_o)\right],
\end{equation}
where $a_o(t)$ is the solution for Eq. (\ref{Happ}) setting $\lambda_o$ to zero, and $\xi(t,\lambda_o)$ is a function such that $\xi\ll 1$, after cancelling second order derivatives of $\xi$, we find
\begin{equation}
    a(t)\approx a_d\left(\alpha H_{\text{dS}}t\right)^{1/\alpha}\left[1-\frac{\lambda_o }{6a_d^3 H_{\text{dS}}^{2+3/\alpha}(\alpha t)^{-1+3/\alpha}}\right].
\end{equation}

Using the slow-roll parameter \cite{Bassett}
\begin{equation}
    \epsilon=-\frac{\dot{H}}{H^2}\approx\alpha-\lambda_o
    \frac{(3-\alpha)(3-2\alpha)(1-\alpha)}{6H_{\text{dS}}^2a^2(\eta)}\eta,
\end{equation}
the Mukhanov-Sasaki equation (\ref{MSEq}), at first order in $\lambda_o$, is just
\begin{equation}\label{DifEq}
    v''_k(\eta)+\left(k^2-\frac{2}{\eta^2}-\frac{3\alpha}{\eta^2}
    +\lambda_o\frac{\left[a_d^{\alpha}H_{\text{dS}}(1-\alpha)\right]^{\frac{2}{1-\alpha}}}{6H_{\text{dS}}^2}|\eta|^{\frac{1+\alpha}{1-\alpha}}\right)v_k(\eta)=0,
\end{equation}
where we have included the relations
\begin{equation}
    \eta\approx-\frac{t\alpha}{(1-\alpha)a(t)}\approx -\frac{1}{(1-\alpha)aH},\qquad \qquad \frac{z''}{z}=(aH)^2(2-\epsilon).
\end{equation}

As expected, the Bunch-Davies vacuum \cite{Mukhanov}
\begin{equation}\label{Vacuum}
    v_k(\eta)=\frac{\e^{-\i k\eta}}{\sqrt{2k}},
\end{equation}
cannot be recovered at $\eta\to-\infty$, since the term due to the interaction between universes (the third one into the parenthesis in Eq. (\ref{DifEq})), dominates, and hence it cannot be used as a good boundary condition. However, since $(aH)\sim \sqrt[3]{\lambda_o}$ during the initial phase, such a size corresponds to the conformal time
\begin{equation}
    \eta_{\text{knee}}\approx-\frac{1}{(1-\alpha)\sqrt[3]{\lambda_o}}\approx-\frac{1}{\sqrt[3]{\lambda_o}},
\end{equation}
for the late time evolution. We will use a solution to Eq. (\ref{DifEq}) such that it behaves like (\ref{Vacuum}) at $\eta_{\text{knee}}$. This approximation is thought to be very good since $\abs{\eta_{\text{knee}}}$ must be very large.

Trying a solution of the form
\begin{equation}\label{Ans}
    v_k(\eta)=v_k^{(0)}(\eta)\e^{f_k(\eta,\lambda_o)},
\end{equation}
where $v_k^{0}(\eta)$ is the solution to Eq. (\ref{DifEq}) when the third term in the parenthesis is vanishing, and $f_k(\eta,\lambda_o)$ is a slow varying function whose value is expected to be small enough in its whole domain. Introducing (\ref{Ans}) into (\ref{DifEq}), and cancelling the terms containing $f''$ and $(f')^2$, we find the solution going like
\begin{equation}
    v_k(\eta)=v_k^{(0)}(\eta)
    \exp\left[-\lambda_o\frac{\left[a_d^{\alpha}H_{\text{dS}}(1-\alpha)\right]^{\frac{2}{1-\alpha}}}{12H_{\text{dS}}^2}\int^{\eta}|\tilde{\eta}|^{\frac{1+\alpha}{1-\alpha}}\frac{v_k^{(0)}(\tilde{\eta})}{{v'_k}^{(0)}(\tilde{\eta})}\d\tilde{\eta}+\Delta_k\right],
\end{equation}
where $\Delta_k$ is a constant of integration. This constant $\Delta_k$ is very close to be purely imaginary, and for the future calculation of the power spectrum the phase to which it contributes is not relevant. The solution $v^{(0)}_k(\eta)$ is well-known as \cite{Bassett} 
\begin{equation}
    v^{(0)}_k(\eta)=\frac{\sqrt{\pi|\eta|}}{2}H^{(1)}_{\mu}\left(k|\eta|\right),\qquad \qquad \mu\approx\frac32+\frac59\alpha,
\end{equation}
where $H_{n}^{(1)}(z)$ is the Hankel function of the first kind (for special functions, see \cite{NIST}), since it fulfils the condition (\ref{Vacuum}) at $\eta\to-\infty$.

The anisotropies of the CMB are expressed via the power spectrum of the curvature perturbations \cite{Mukhanov}
\begin{equation}\label{Power}
    P_{\mathcal{R}}(k)=\frac{k^3}{2\pi^2}\frac{|v_k|^2}{z^2}.
\end{equation}
One expects to adjust the spectrum to a power law like
\begin{equation}\label{FixPower}
    P_{\mathcal{R}}(k)=A_s\left(\frac{k}{k_{\ast}}\right)^{n_s-1},
\end{equation}
when $aH\sim k$, or equivalently, $k|\eta|\sim1$, where $A_s$, and $n_s$ depend on the given value $k_{\ast}$ set by the experiment. From the Planck analysis \cite{Planck}, where they considered $k_{\ast}=0.05$ Mpc$^{-1}$, it was obtained $A_s=(2.105\pm0.030)\times10^{-9}$, and $n_s=(0.9665\pm0.0038)$.
That way, the power spectrum is found to have the known form \cite{Interacting,Morais,Bassett}
\begin{equation}
    P_{\mathcal{R}}(k\lesssim aH)\approx \left[(1-\alpha)\frac{\Gamma(\nu)}{\Gamma(3/2)}\frac{H}{2\pi}\right]^2\left(\frac{k|\eta|}{2}\right)^{-2\alpha}\left|\e^{f_k(\eta,\lambda_o)}\right|^2,
\end{equation}
which is found, in our case, at first order in $\epsilon\sim\alpha$, and at the moment when the modes cross the horizon $k_{\ast}\approx a_{\ast}H_{\text{dS}}$, to be like
\begin{equation}
    P_{\mathcal{R}}(k\lesssim aH)\approx \frac{1}{2^{-2\alpha}}
    \left\{\left[1-\alpha+\Psi\left(\frac32\right)\alpha\right]\frac{H_{\text{dS}}}{2\pi}\right\}^2\left(\frac{k}{a_{\ast}H_{\text{dS}}}\right)^{-2\alpha}\left|\e^{f_k(\eta,\lambda_o)}\right|^2,
\end{equation}
where $\Psi(z):=\Gamma'(z)/\Gamma(z)$ is the digamma function. The absolute value of the exponential can be considered very close to unity such that we can derive, comparing with the fit (\ref{FixPower}), the values
\begin{gather}
    \alpha=0.0168\pm0.0019, \qquad \qquad  a_d\equiv a_{\ast}=(9.052\pm0.067)\cdot10^{-56}, \\ H_{\text{dS}}=(2.896\pm0.021)\cdot10^{-4}=(5.372\pm0.039) \cdot10^{39}\,\text{s}^{-1},
\end{gather}
which we can compare with the actual values
\begin{equation}
    a_0=1,\qquad \qquad 
    H_0\approx1.2\cdot 10^{-61}.
\end{equation}

\begin{figure}[t]
    \centering
    \includegraphics[width=\textwidth]{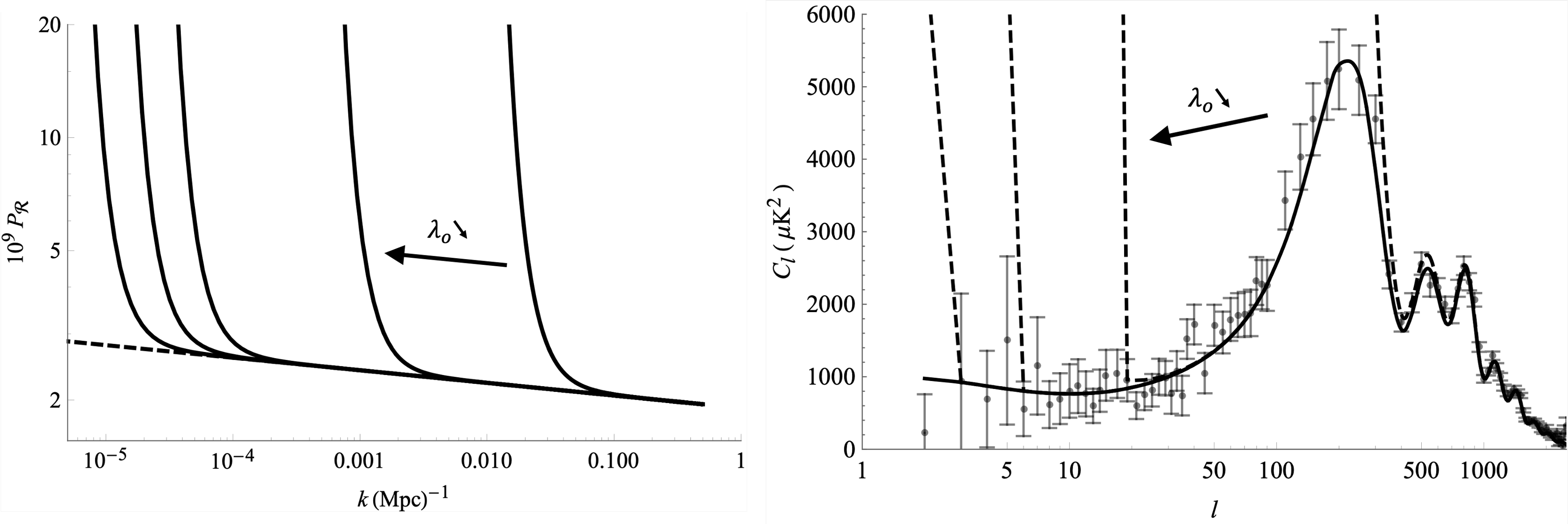}
    \caption{{\it Left panel:} The power spectrum for different values of $\lambda_o$. The dashed line represents the spectrum (\ref{FixPower}) given by Planck data. The smaller the value of $\lambda_o$, the spectrum is more similar to the standard, as expected. The effect of the multiverse interaction is visible only for the smaller modes $k$. {\it Right panel:} The normalized angular power spectrum for different values of $\lambda_o$. The solid line represents the standard spectrum. The data points and the error bars come from the Planck measurement. The dashed lines are the spectrum found for different $\lambda_o$. For $\lambda_o=10^{-56}$, the dashed line is overlapped by the solid line. Whence, the smaller the $\lambda_o$, the lesser the affect onto the multipoles. For both figures, the considered values for the coupling constant $\lambda_o$ are: $10^{-46}$, $10^{-50}$, $10^{-54}$, $10^{-55}$, $10^{-56}$. }
    \label{Fig1}
\end{figure}

The power spectrum (\ref{Power}) is used then to find the angular power spectrum, given in terms of the coefficients
\begin{equation}
    C_l=2T_0^2l(l+1)\int_0^{\infty} \frac{\d k}{k} P_{\mathcal{R}}(k)\Delta_l^2(k),
\end{equation}
where $T_0=(2.72548\pm0.00057)$ K is the temperature of the CMB \cite{Fixsen2009}, and \textsc{camb} has been used to obtain the transfer functions $\Delta_l(k)$. For us it depends on $\lambda_o$. The power spectrum (\ref{Power}) and the angular power spectrum are shown in Fig. \ref{Fig1}. 

As expected, the interaction with our hypothetical partner universe affects only the smaller values of $k$ and the multipoles. The lighter the interaction, the more suppressed the effects. Since the spectra differ too fast from the standard one, we can easily constraint the value of the coupling constant $\lambda_o$, for a kind of coupling function as (\ref{Coupling}), to be
\begin{equation}
    \lambda_o \lesssim \mathcal{O}(10^{-56}).
\end{equation}

Coming back to a more general coupling function like in (\ref{Couplingq}), the spectrum for a certain family is expected to be very similar as the one in Fig. \ref{Fig1}. Such family is analogous to the one we analysed, but the condition is that $4-q+m-n>0$ in order to keep the vacuum stability assuming $\dot{a}>0$ and taking $m$ even, and $2-q+m-n=0$, which automatically satisfies the first condition.  Within the family we find the choice $m=2$, $n=4$ and $q=0$, for which the Friedmann equation (\ref{FE}) is
\begin{equation}
    H^2=H_{\text{dS}}^2\left[\left(\frac{a_d}{a}\right)^{2\alpha}+\frac{\lambda_o}{H_{\text{dS}}^2a^2\ddot{a}^2}\right].
\end{equation}

The scale factor goes like $t^{3/2}$ close to the initial singularity, and taking $\lambda_o$ as a perturbation for late times, the scale factor is then evolving like $t^{1/\alpha}$. The transition between both states is expected to be smooth, like if a single scalar field whose barotropic parameter varies from $\omega=-5/9$ to $\omega\gtrsim-1$ monotonically fills the universe. It implies that the power spectrum for the modes crossing the horizon $k\sim aH$, during the early states of the universe, when the interaction dominates, is just enlarged. Those modes are the smallest ones. It will not change significantly the results in Fig. (\ref{Fig1}) since the slopes are quite steep already for small $k$.

\section{A Short Comment About an Interuniversal Contribution to the Dark Energy of the Universe}\label{SDE}

Here we shortly present a way for the multiverse interaction to contribute to the dark energy of the universe via its entanglement. The Friedmann equation (\ref{FE}) rules the dynamics of the universe with mode $K=1$ for a given interaction $\lambda(a,H)$ as we have reasoned previously.  Now, let us consider the special form of the coupling function
\begin{equation}\label{ExLambda}
\lambda^2(a,H)=\frac{\lambda_o}{2}a^4f(H)H^2,
\end{equation}
where $f(H)$ is an explicit and well-behaved function of $H$, and $\lambda_o$ is a positive constant. In order to be sure that the interaction term diverges when $H$ vanishes, we require $f(H)H^2$ to diverge. The Friedmann equation (\ref{FE}), without any field, yields the trivial result
\begin{equation}
    H=f^{-1}(\lambda_o^{-1})=\text{constant}.
\end{equation}
The dynamics is then equivalent to the one of the de Sitter space
\begin{equation}
    a(t)=a_o\exp\left(Ht\right).
\end{equation}

Even if the Friedmann equation (\ref{FE}) has been obtained without assuming that such coupling function was small, one  needs it to be so small since the probability condition (\ref{Cond}), which is of the form $\mathcal{P}\sim\exp\{-[\lambda_oH^2f(H)]^{-3}\}$, prevents $\lambda_o$ to be very large in order to keep a small initial probability. Besides, the probability condition tells us that not considering any kind of matter produces a constant probability along the entire evolution of the universe. There is no other chance than having some kind of matter contributing to the dynamics, so that the probability decreases leading to the recovery of the anisotropies of the CMB.

\section{Conclusions}
\label{SConc}

The multiverse scenario derived from the third quantization in quantum cosmology has been considered. In this scenario, the universes behave like particles of the standard quantum field theory. We have considered the interaction of a pair of universes (our universe and our twin anti-universe), whose entanglement entropy was found to diverge at the critical points of the classical evolution \cite{Critical}. It has motivated us to consider the coupling function like (\ref{Coupling}) or a more general one like (\ref{Couplingq}), which both depend on an arbitrary coupling constant $\lambda_o$ which is to be determined and on the Hubble parameter. We have assumed that the functions must fulfil the condition (\ref{Limit}) (coupling function diverges when the Hubble parameter vanishes) and the condition (\ref{Cond}) (the probability is supressed at large values of the scale factor). For the function (\ref{Coupling}), we have found the modified Friedmann equation (\ref{HLambda}) while taking into account the interaction of the pair of universes.

For a certain family of solutions, we have found the modifications of the imprints on the CMB power spectrum. The plots have been given in Fig. \ref{Fig1} in which one can see that the interaction with our hypothetical partner enlarges the spectra for small $k$ and the small monopoles $l$.  The constraint on the interaction coupling constant which is in agreement with Planck observations \cite{PlanckSpectrum} has been found to be such as $\lambda_o\lesssim\mathcal{O}(10^{-56})$.

It is the future problem to fix the parameters of the coupling function and its specific form. However, for the set of solutions fulfilling the condition $2-q-n=0$, where $n>0$, the results presented in the paper are approximately equivalent. Once those parameters are known, then one can extrapolate our conclusions and say that the CMB angular power spectrum is insensitive to some small interaction term. The need for more observational experiments digging into the very early stages of our universe, like the search for primordial gravitational waves or the cosmic neutrino background, would be of interest in order to falsify the existence of a twin anti-universe or some other entangled universes. From what we have found, we can only constrain the strength of the interaction, which is considerably small. 

A short comment has also been given for two empty universes whose interaction was defined by a very special coupling function (\ref{ExLambda}). The dynamics of such universes is similar to the de Sitter universe. It could mimic the effect of a cosmological constant at early times. Nevertheless, in order to recover the spectra, one needs some kind of matter to be introduced to such a simple scenario.

\acknowledgments

SBB would like to thank Michele Liguori and Eleonora Di Valentino for providing useful hints to calculate the CMB angular power spectrum and Mar Bastero Gil and Fabian Wagner for some useful discussions and coffees. The work of SBB was supported by the Polish National Research and Development Center (NCBR) project UNIWERSYTET 2.0. STREFA KARIERY, POWR.03.05.00-00-Z064/17-00.

\bibliography{Bib.bib}

%merlin.mbs apsrev4-1.bst 2010-07-25 4.21a (PWD, AO, DPC) hacked
%Control: key (0)
%Control: author (8) initials jnrlst
%Control: editor formatted (1) identically to author
%Control: production of article title (-1) disabled
%Control: page (0) single
%Control: year (1) truncated
%Control: production of eprint (0) enabled
\begin{thebibliography}{42}%
\makeatletter
\providecommand \@ifxundefined [1]{%
 \@ifx{#1\undefined}
}%
\providecommand \@ifnum [1]{%
 \ifnum #1\expandafter \@firstoftwo
 \else \expandafter \@secondoftwo
 \fi
}%
\providecommand \@ifx [1]{%
 \ifx #1\expandafter \@firstoftwo
 \else \expandafter \@secondoftwo
 \fi
}%
\providecommand \natexlab [1]{#1}%
\providecommand \enquote  [1]{``#1''}%
\providecommand \bibnamefont  [1]{#1}%
\providecommand \bibfnamefont [1]{#1}%
\providecommand \citenamefont [1]{#1}%
\providecommand \href@noop [0]{\@secondoftwo}%
\providecommand \href [0]{\begingroup \@sanitize@url \@href}%
\providecommand \@href[1]{\@@startlink{#1}\@@href}%
\providecommand \@@href[1]{\endgroup#1\@@endlink}%
\providecommand \@sanitize@url [0]{\catcode `\\12\catcode `\$12\catcode
  `\&12\catcode `\#12\catcode `\^12\catcode `\_12\catcode `\%12\relax}%
\providecommand \@@startlink[1]{}%
\providecommand \@@endlink[0]{}%
\providecommand \url  [0]{\begingroup\@sanitize@url \@url }%
\providecommand \@url [1]{\endgroup\@href {#1}{\urlprefix }}%
\providecommand \urlprefix  [0]{URL }%
\providecommand \Eprint [0]{\href }%
\providecommand \doibase [0]{http://dx.doi.org/}%
\providecommand \selectlanguage [0]{\@gobble}%
\providecommand \bibinfo  [0]{\@secondoftwo}%
\providecommand \bibfield  [0]{\@secondoftwo}%
\providecommand \translation [1]{[#1]}%
\providecommand \BibitemOpen [0]{}%
\providecommand \bibitemStop [0]{}%
\providecommand \bibitemNoStop [0]{.\EOS\space}%
\providecommand \EOS [0]{\spacefactor3000\relax}%
\providecommand \BibitemShut  [1]{\csname bibitem#1\endcsname}%
\let\auto@bib@innerbib\@empty
%</preamble>
\bibitem [{\citenamefont {Langhoff}\ \emph {et~al.}(2021)\citenamefont
  {Langhoff}, \citenamefont {Murdia},\ and\ \citenamefont {Nomura}}]{Multi1}%
  \BibitemOpen
  \bibfield  {author} {\bibinfo {author} {\bibfnamefont {K.}~\bibnamefont
  {Langhoff}}, \bibinfo {author} {\bibfnamefont {C.}~\bibnamefont {Murdia}}, \
  and\ \bibinfo {author} {\bibfnamefont {Y.}~\bibnamefont {Nomura}},\ }\href
  {\doibase 10.1103/physrevd.104.086007} {\bibfield  {journal} {\bibinfo
  {journal} {Physical Review D}\ }\textbf {\bibinfo {volume} {104}},\ \bibinfo
  {pages} {086007} (\bibinfo {year} {2021})}\BibitemShut {NoStop}%
\bibitem [{\citenamefont {Aguilar-Gutierrez}\ \emph {et~al.}(2021)\citenamefont
  {Aguilar-Gutierrez}, \citenamefont {Chatwin-Davies}, \citenamefont {Hertog},
  \citenamefont {Pinzani-Fokeeva},\ and\ \citenamefont {Robinson}}]{Multi2}%
  \BibitemOpen
  \bibfield  {author} {\bibinfo {author} {\bibfnamefont {S.~E.}\ \bibnamefont
  {Aguilar-Gutierrez}}, \bibinfo {author} {\bibfnamefont {A.}~\bibnamefont
  {Chatwin-Davies}}, \bibinfo {author} {\bibfnamefont {T.}~\bibnamefont
  {Hertog}}, \bibinfo {author} {\bibfnamefont {N.}~\bibnamefont
  {Pinzani-Fokeeva}}, \ and\ \bibinfo {author} {\bibfnamefont {B.}~\bibnamefont
  {Robinson}},\ }\href {\doibase 10.1007/jhep11(2021)212} {\bibfield  {journal}
  {\bibinfo  {journal} {Journal of High Energy Physics}\ }\textbf {\bibinfo
  {volume} {2021}},\ \bibinfo {pages} {212} (\bibinfo {year}
  {2021})}\BibitemShut {NoStop}%
\bibitem [{\citenamefont {Alonso-Serrano}\ and\ \citenamefont
  {Jannes}(2019)}]{Multi3}%
  \BibitemOpen
  \bibfield  {author} {\bibinfo {author} {\bibfnamefont {A.}~\bibnamefont
  {Alonso-Serrano}}\ and\ \bibinfo {author} {\bibfnamefont {G.}~\bibnamefont
  {Jannes}},\ }\href {\doibase 10.3390/universe5100212} {\bibfield  {journal}
  {\bibinfo  {journal} {Universe}\ }\textbf {\bibinfo {volume} {5}},\ \bibinfo
  {pages} {212} (\bibinfo {year} {2019})}\BibitemShut {NoStop}%
\bibitem [{\citenamefont {D\c{a}browski}(2019)}]{universe2019}%
  \BibitemOpen
  \bibfield  {author} {\bibinfo {author} {\bibfnamefont {M.~P.}\ \bibnamefont
  {D\c{a}browski}},\ }\href {\doibase 10.3390/universe5070172} {\bibfield
  {journal} {\bibinfo  {journal} {Universe}\ }\textbf {\bibinfo {volume} {5}},\
  \bibinfo {pages} {172} (\bibinfo {year} {2019})}\BibitemShut {NoStop}%
\bibitem [{\citenamefont {Linde}\ and\ \citenamefont
  {Vanchurin}(2010)}]{Multi4}%
  \BibitemOpen
  \bibfield  {author} {\bibinfo {author} {\bibfnamefont {A.}~\bibnamefont
  {Linde}}\ and\ \bibinfo {author} {\bibfnamefont {V.}~\bibnamefont
  {Vanchurin}},\ }\href {\doibase 10.1103/physrevd.81.083525} {\bibfield
  {journal} {\bibinfo  {journal} {Physical Review D}\ }\textbf {\bibinfo
  {volume} {81}},\ \bibinfo {pages} {083525} (\bibinfo {year}
  {2010})}\BibitemShut {NoStop}%
\bibitem [{\citenamefont {Hartle}(2021)}]{Multi5}%
  \BibitemOpen
  \bibfield  {author} {\bibinfo {author} {\bibfnamefont {J.~B.}\ \bibnamefont
  {Hartle}},\ }\href {\doibase 10.1142/11716} {\emph {\bibinfo {title} {The
  Quantum Universe}}}\ (\bibinfo  {publisher} {WORLD SCIENTIFIC},\ \bibinfo
  {year} {2021})\ \Eprint
  {http://arxiv.org/abs/https://www.worldscientific.com/doi/pdf/10.1142/11716}
  {https://www.worldscientific.com/doi/pdf/10.1142/11716} \BibitemShut
  {NoStop}%
\bibitem [{\citenamefont {Garriga}\ \emph {et~al.}(2016)\citenamefont
  {Garriga}, \citenamefont {Vilenkin},\ and\ \citenamefont {Zhang}}]{Multi6}%
  \BibitemOpen
  \bibfield  {author} {\bibinfo {author} {\bibfnamefont {J.}~\bibnamefont
  {Garriga}}, \bibinfo {author} {\bibfnamefont {A.}~\bibnamefont {Vilenkin}}, \
  and\ \bibinfo {author} {\bibfnamefont {J.}~\bibnamefont {Zhang}},\ }\href
  {\doibase 10.1088/1475-7516/2016/02/064} {\bibfield  {journal} {\bibinfo
  {journal} {Journal of Cosmology and Astroparticle Physics}\ }\textbf
  {\bibinfo {volume} {2016}},\ \bibinfo {pages} {064} (\bibinfo {year}
  {2016})}\BibitemShut {NoStop}%
\bibitem [{\citenamefont {Mersini-Houghton}(2017)}]{Multi7}%
  \BibitemOpen
  \bibfield  {author} {\bibinfo {author} {\bibfnamefont {L.}~\bibnamefont
  {Mersini-Houghton}},\ }\href {\doibase 10.1088/1361-6382/34/4/047001}
  {\bibfield  {journal} {\bibinfo  {journal} {Classical and Quantum Gravity}\
  }\textbf {\bibinfo {volume} {34}},\ \bibinfo {pages} {047001} (\bibinfo
  {year} {2017})}\BibitemShut {NoStop}%
\bibitem [{\citenamefont {Vilenkin}(2014)}]{Multi8}%
  \BibitemOpen
  \bibfield  {author} {\bibinfo {author} {\bibfnamefont {A.}~\bibnamefont
  {Vilenkin}},\ }\href {\doibase 10.1088/1475-7516/2014/05/005} {\bibfield
  {journal} {\bibinfo  {journal} {Journal of Cosmology and Astroparticle
  Physics}\ }\textbf {\bibinfo {volume} {2014}},\ \bibinfo {pages} {005}
  (\bibinfo {year} {2014})}\BibitemShut {NoStop}%
\bibitem [{\citenamefont {Balcerzak}\ and\ \citenamefont
  {Lisaj}(2021)}]{BalcerzakBell}%
  \BibitemOpen
  \bibfield  {author} {\bibinfo {author} {\bibfnamefont {A.}~\bibnamefont
  {Balcerzak}}\ and\ \bibinfo {author} {\bibfnamefont {M.}~\bibnamefont
  {Lisaj}},\ }\href@noop {} {\  (\bibinfo {year} {2021})},\ \Eprint
  {http://arxiv.org/abs/2112.14166} {arXiv:2112.14166 [gr-qc]} \BibitemShut
  {NoStop}%
\bibitem [{\citenamefont {Bouhmadi-L{\'{o}}pez}\ \emph
  {et~al.}(2019)\citenamefont {Bouhmadi-L{\'{o}}pez}, \citenamefont {Kr\"amer},
  \citenamefont {Morais},\ and\ \citenamefont
  {Robles-P{\'{e}}rez}}]{Interacting}%
  \BibitemOpen
  \bibfield  {author} {\bibinfo {author} {\bibfnamefont {M.}~\bibnamefont
  {Bouhmadi-L{\'{o}}pez}}, \bibinfo {author} {\bibfnamefont {M.}~\bibnamefont
  {Kr\"amer}}, \bibinfo {author} {\bibfnamefont {J.}~\bibnamefont {Morais}}, \
  and\ \bibinfo {author} {\bibfnamefont {S.}~\bibnamefont
  {Robles-P{\'{e}}rez}},\ }\href {\doibase 10.1088/1475-7516/2019/02/057}
  {\bibfield  {journal} {\bibinfo  {journal} {Journal of Cosmology and
  Astroparticle Physics}\ }\textbf {\bibinfo {volume} {2019}},\ \bibinfo
  {pages} {057} (\bibinfo {year} {2019})}\BibitemShut {NoStop}%
\bibitem [{\citenamefont {Tegmark}(2004)}]{Tegmark}%
  \BibitemOpen
  \bibfield  {author} {\bibinfo {author} {\bibfnamefont {M.}~\bibnamefont
  {Tegmark}},\ }in\ \href@noop {} {\emph {\bibinfo {booktitle} {Science and
  Ultimate Reality: Quantum Theory, Cosmology, and Complexity}}},\ \bibinfo
  {editor} {edited by\ \bibinfo {editor} {\bibfnamefont {J.~D.}\ \bibnamefont
  {Barrow}}, \bibinfo {editor} {\bibfnamefont {P.~C.~W.}\ \bibnamefont
  {Davies}}, \ and\ \bibinfo {editor} {\bibfnamefont {C.~L.}\ \bibnamefont
  {Harper}}}\ (\bibinfo  {publisher} {Cambridge University Press},\ \bibinfo
  {year} {2004})\ p.\ \bibinfo {pages} {459}\BibitemShut {NoStop}%
\bibitem [{\citenamefont {Caderni}\ and\ \citenamefont
  {Martellini}(1984)}]{3Q1}%
  \BibitemOpen
  \bibfield  {author} {\bibinfo {author} {\bibfnamefont {N.}~\bibnamefont
  {Caderni}}\ and\ \bibinfo {author} {\bibfnamefont {M.}~\bibnamefont
  {Martellini}},\ }\href@noop {} {\bibfield  {journal} {\bibinfo  {journal}
  {International Journal of Theoretical Physics}\ }\textbf {\bibinfo {volume}
  {23}},\ \bibinfo {pages} {233} (\bibinfo {year} {1984})}\BibitemShut
  {NoStop}%
\bibitem [{\citenamefont {McGuigan}(1988)}]{3Q2}%
  \BibitemOpen
  \bibfield  {author} {\bibinfo {author} {\bibfnamefont {M.}~\bibnamefont
  {McGuigan}},\ }\href {\doibase 10.1103/PhysRevD.38.3031} {\bibfield
  {journal} {\bibinfo  {journal} {Phys. Rev. D}\ }\textbf {\bibinfo {volume}
  {38}},\ \bibinfo {pages} {3031} (\bibinfo {year} {1988})}\BibitemShut
  {NoStop}%
\bibitem [{\citenamefont {Giddings}\ and\ \citenamefont
  {Strominger}(1989)}]{Giddings1988}%
  \BibitemOpen
  \bibfield  {author} {\bibinfo {author} {\bibfnamefont {S.~B.}\ \bibnamefont
  {Giddings}}\ and\ \bibinfo {author} {\bibfnamefont {A.}~\bibnamefont
  {Strominger}},\ }\href {\doibase 10.1016/0550-3213(89)90353-2} {\bibfield
  {journal} {\bibinfo  {journal} {Nucl. Phys. B}\ }\textbf {\bibinfo {volume}
  {321}},\ \bibinfo {pages} {481} (\bibinfo {year} {1989})}\BibitemShut
  {NoStop}%
\bibitem [{\citenamefont {Robles-P{\'e}rez}(2021)}]{3QSalva}%
  \BibitemOpen
  \bibfield  {author} {\bibinfo {author} {\bibfnamefont {S.~J.}\ \bibnamefont
  {Robles-P{\'e}rez}},\ }\href {https://www.mdpi.com/2218-1997/7/11/404}
  {\bibfield  {journal} {\bibinfo  {journal} {Universe}\ }\textbf {\bibinfo
  {volume} {7}} (\bibinfo {year} {2021})}\BibitemShut {NoStop}%
\bibitem [{\citenamefont {DeWitt}(1967)}]{DeWitt1}%
  \BibitemOpen
  \bibfield  {author} {\bibinfo {author} {\bibfnamefont {B.~S.}\ \bibnamefont
  {DeWitt}},\ }\href {\doibase 10.1103/PhysRev.160.1113} {\bibfield  {journal}
  {\bibinfo  {journal} {Phys. Rev.}\ }\textbf {\bibinfo {volume} {160}},\
  \bibinfo {pages} {1113} (\bibinfo {year} {1967})}\BibitemShut {NoStop}%
\bibitem [{\citenamefont {Kiefer}(2007)}]{KieferBook}%
  \BibitemOpen
  \bibfield  {author} {\bibinfo {author} {\bibfnamefont {C.}~\bibnamefont
  {Kiefer}},\ }\href@noop {} {\emph {\bibinfo {title} {{Quantum Gravity}}}},\
  \bibinfo {edition} {2nd}\ ed.\ (\bibinfo  {publisher} {Oxford University
  Press},\ \bibinfo {address} {New York},\ \bibinfo {year} {2007})\BibitemShut
  {NoStop}%
\bibitem [{\citenamefont {Robles-P{\'e}rez}\ and\ \citenamefont
  {Gonz{\'a}lez-D{\'i}az}(2010)}]{Babies1}%
  \BibitemOpen
  \bibfield  {author} {\bibinfo {author} {\bibfnamefont {S.}~\bibnamefont
  {Robles-P{\'e}rez}}\ and\ \bibinfo {author} {\bibfnamefont {P.~F.}\
  \bibnamefont {Gonz{\'a}lez-D{\'i}az}},\ }\href {\doibase
  10.1103/physrevd.81.083529} {\bibfield  {journal} {\bibinfo  {journal}
  {Physical Review D}\ }\textbf {\bibinfo {volume} {81}} (\bibinfo {year}
  {2010}),\ 10.1103/physrevd.81.083529}\BibitemShut {NoStop}%
\bibitem [{\citenamefont {Robles-P{\'e}rez}(2020)}]{Babies2}%
  \BibitemOpen
  \bibfield  {author} {\bibinfo {author} {\bibfnamefont {S.~J.}\ \bibnamefont
  {Robles-P{\'e}rez}},\ }\href {\doibase 10.5506/APhysPolBSupp.13.325}
  {\bibfield  {journal} {\bibinfo  {journal} {Acta Phys. Polon. Supp.}\
  }\textbf {\bibinfo {volume} {13}},\ \bibinfo {pages} {325} (\bibinfo {year}
  {2020})},\ \Eprint {http://arxiv.org/abs/2002.09863} {arXiv:2002.09863
  [gr-qc]} \BibitemShut {NoStop}%
\bibitem [{\citenamefont {Schwinger}(1951)}]{SchwingerEffect}%
  \BibitemOpen
  \bibfield  {author} {\bibinfo {author} {\bibfnamefont {J.}~\bibnamefont
  {Schwinger}},\ }\href {\doibase 10.1103/PhysRev.82.664} {\bibfield  {journal}
  {\bibinfo  {journal} {Phys. Rev.}\ }\textbf {\bibinfo {volume} {82}},\
  \bibinfo {pages} {664} (\bibinfo {year} {1951})}\BibitemShut {NoStop}%
\bibitem [{\citenamefont {Robles-P{\'e}rez}\ \emph {et~al.}(2017)\citenamefont
  {Robles-P{\'e}rez}, \citenamefont {Balcerzak}, \citenamefont
  {D\c{a}browski},\ and\ \citenamefont {Kr\"amer}}]{Cyclic}%
  \BibitemOpen
  \bibfield  {author} {\bibinfo {author} {\bibfnamefont {S.}~\bibnamefont
  {Robles-P{\'e}rez}}, \bibinfo {author} {\bibfnamefont {A.}~\bibnamefont
  {Balcerzak}}, \bibinfo {author} {\bibfnamefont {M.~P.}\ \bibnamefont
  {D\c{a}browski}}, \ and\ \bibinfo {author} {\bibfnamefont {M.}~\bibnamefont
  {Kr\"amer}},\ }\href {\doibase 10.1103/physrevd.95.083505} {\bibfield
  {journal} {\bibinfo  {journal} {Physical Review D}\ }\textbf {\bibinfo
  {volume} {95}} (\bibinfo {year} {2017}),\
  10.1103/physrevd.95.083505}\BibitemShut {NoStop}%
\bibitem [{\citenamefont {Balcerzak}\ \emph {et~al.}(2021)\citenamefont
  {Balcerzak}, \citenamefont {Barroso-Bellido}, \citenamefont {D\k{a}browski},\
  and\ \citenamefont {Robles-P\'erez}}]{Critical}%
  \BibitemOpen
  \bibfield  {author} {\bibinfo {author} {\bibfnamefont {A.}~\bibnamefont
  {Balcerzak}}, \bibinfo {author} {\bibfnamefont {S.}~\bibnamefont
  {Barroso-Bellido}}, \bibinfo {author} {\bibfnamefont {M.~P.}\ \bibnamefont
  {D\k{a}browski}}, \ and\ \bibinfo {author} {\bibfnamefont {S.}~\bibnamefont
  {Robles-P\'erez}},\ }\href {\doibase 10.1103/PhysRevD.103.043507} {\bibfield
  {journal} {\bibinfo  {journal} {Phys. Rev. D}\ }\textbf {\bibinfo {volume}
  {103}},\ \bibinfo {pages} {043507} (\bibinfo {year} {2021})}\BibitemShut
  {NoStop}%
\bibitem [{\citenamefont {Bellido}(2021)}]{SBB2}%
  \BibitemOpen
  \bibfield  {author} {\bibinfo {author} {\bibfnamefont {S.~B.}\ \bibnamefont
  {Bellido}},\ }\href {\doibase 10.1103/PhysRevD.104.106009} {\bibfield
  {journal} {\bibinfo  {journal} {Phys. Rev. D}\ }\textbf {\bibinfo {volume}
  {104}},\ \bibinfo {pages} {106009} (\bibinfo {year} {2021})}\BibitemShut
  {NoStop}%
\bibitem [{\citenamefont {Kiefer}\ and\ \citenamefont {Zeh}(1995)}]{kiefer95}%
  \BibitemOpen
  \bibfield  {author} {\bibinfo {author} {\bibfnamefont {C.}~\bibnamefont
  {Kiefer}}\ and\ \bibinfo {author} {\bibfnamefont {H.~D.}\ \bibnamefont
  {Zeh}},\ }\href {\doibase 10.1103/physrevd.51.4145} {\bibfield  {journal}
  {\bibinfo  {journal} {Physical Review D}\ }\textbf {\bibinfo {volume} {51}},\
  \bibinfo {pages} {4145} (\bibinfo {year} {1995})}\BibitemShut {NoStop}%
\bibitem [{\citenamefont {D\c{a}browski}\ and\ \citenamefont
  {Larsen}(1995)}]{MPD95}%
  \BibitemOpen
  \bibfield  {author} {\bibinfo {author} {\bibfnamefont {M.}~\bibnamefont
  {D\c{a}browski}}\ and\ \bibinfo {author} {\bibfnamefont {A.}~\bibnamefont
  {Larsen}},\ }\href {\doibase 10.1103/physrevd.52.3424} {\bibfield  {journal}
  {\bibinfo  {journal} {Physical Review D}\ }\textbf {\bibinfo {volume} {52}},\
  \bibinfo {pages} {3424–3431} (\bibinfo {year} {1995})}\BibitemShut
  {NoStop}%
\bibitem [{\citenamefont {Robles-P{\'e}rez}\ \emph {et~al.}(2016)\citenamefont
  {Robles-P{\'e}rez}, \citenamefont {Alonso-Serrano}, \citenamefont {Bastos},\
  and\ \citenamefont {Bertolami}}]{Vacuum}%
  \BibitemOpen
  \bibfield  {author} {\bibinfo {author} {\bibfnamefont {S.}~\bibnamefont
  {Robles-P{\'e}rez}}, \bibinfo {author} {\bibfnamefont {A.}~\bibnamefont
  {Alonso-Serrano}}, \bibinfo {author} {\bibfnamefont {C.}~\bibnamefont
  {Bastos}}, \ and\ \bibinfo {author} {\bibfnamefont {O.}~\bibnamefont
  {Bertolami}},\ }\href {\doibase
  https://doi.org/10.1016/j.physletb.2016.05.091} {\bibfield  {journal}
  {\bibinfo  {journal} {Physics Letters B}\ }\textbf {\bibinfo {volume}
  {759}},\ \bibinfo {pages} {328} (\bibinfo {year} {2016})}\BibitemShut
  {NoStop}%
\bibitem [{\citenamefont {Morais}\ \emph {et~al.}(2018)\citenamefont {Morais},
  \citenamefont {Bouhmadi-L{\'o}pez}, \citenamefont {Kr\"amer},\ and\
  \citenamefont {Robles-P{\'e}rez}}]{Morais}%
  \BibitemOpen
  \bibfield  {author} {\bibinfo {author} {\bibfnamefont {J.}~\bibnamefont
  {Morais}}, \bibinfo {author} {\bibfnamefont {M.}~\bibnamefont
  {Bouhmadi-L{\'o}pez}}, \bibinfo {author} {\bibfnamefont {M.}~\bibnamefont
  {Kr\"amer}}, \ and\ \bibinfo {author} {\bibfnamefont {S.}~\bibnamefont
  {Robles-P{\'e}rez}},\ }\href {\doibase 10.1140/epjc/s10052-018-5698-z}
  {\bibfield  {journal} {\bibinfo  {journal} {The European Physical Journal C}\
  }\textbf {\bibinfo {volume} {78}},\ \bibinfo {pages} {5698} (\bibinfo {year}
  {2018})}\BibitemShut {NoStop}%
\bibitem [{\citenamefont {Weinberg}(2008)}]{Weinberg}%
  \BibitemOpen
  \bibfield  {author} {\bibinfo {author} {\bibfnamefont {S.}~\bibnamefont
  {Weinberg}},\ }\href@noop {} {\emph {\bibinfo {title} {Cosmology}}}\
  (\bibinfo  {publisher} {Oxford University Press},\ \bibinfo {year}
  {2008})\BibitemShut {NoStop}%
\bibitem [{\citenamefont {Aghanim}\ \emph
  {et~al.}(2020{\natexlab{a}})\citenamefont {Aghanim}, \citenamefont {Akrami},
  \citenamefont {Ashdown}, \citenamefont {Aumont}, \citenamefont {Baccigalupi},
  \citenamefont {Ballardini}, \citenamefont {Banday}, \citenamefont {Barreiro},
  \citenamefont {Bartolo},\ and\ \citenamefont {et~al.}}]{PlanckSpectrum}%
  \BibitemOpen
  \bibfield  {author} {\bibinfo {author} {\bibfnamefont {N.}~\bibnamefont
  {Aghanim}}, \bibinfo {author} {\bibfnamefont {Y.}~\bibnamefont {Akrami}},
  \bibinfo {author} {\bibfnamefont {M.}~\bibnamefont {Ashdown}}, \bibinfo
  {author} {\bibfnamefont {J.}~\bibnamefont {Aumont}}, \bibinfo {author}
  {\bibfnamefont {C.}~\bibnamefont {Baccigalupi}}, \bibinfo {author}
  {\bibfnamefont {M.}~\bibnamefont {Ballardini}}, \bibinfo {author}
  {\bibfnamefont {A.~J.}\ \bibnamefont {Banday}}, \bibinfo {author}
  {\bibfnamefont {R.~B.}\ \bibnamefont {Barreiro}}, \bibinfo {author}
  {\bibfnamefont {N.}~\bibnamefont {Bartolo}}, \ and\ \bibinfo {author}
  {\bibnamefont {et~al.}},\ }\href {\doibase 10.1051/0004-6361/201936386}
  {\bibfield  {journal} {\bibinfo  {journal} {Astronomy \& Astrophysics}\
  }\textbf {\bibinfo {volume} {641}},\ \bibinfo {pages} {A5} (\bibinfo {year}
  {2020}{\natexlab{a}})}\BibitemShut {NoStop}%
\bibitem [{\citenamefont {D\c{a}browski}(2014)}]{Krakow2014}%
  \BibitemOpen
  \bibfield  {author} {\bibinfo {author} {\bibfnamefont {M.}~\bibnamefont
  {D\c{a}browski}},\ }\enquote {\bibinfo {title} {Are singularities the limits
  of cosmology?}}\ in\ \href@noop {} {\emph {\bibinfo {booktitle} {Mathematical
  Structures of the Universe}}},\ \bibinfo {editor} {edited by\ \bibinfo
  {editor} {\bibfnamefont {M.}~\bibnamefont {Eckstein}}, \bibinfo {editor}
  {\bibfnamefont {M.}~\bibnamefont {Heller}}, \ and\ \bibinfo {editor}
  {\bibfnamefont {S.}~\bibnamefont {Szybka}}}\ (\bibinfo  {publisher}
  {Copernicus Center Press, Krak\'ow},\ \bibinfo {year} {2014})\ Chap.\
  \bibinfo {chapter} {Are Singularities the Limits of Cosmology?}, p.~\bibinfo
  {pages} {99}\BibitemShut {NoStop}%
\bibitem [{\citenamefont {D\c{a}browski}\ and\ \citenamefont
  {Marosek}(2018)}]{GRG2018}%
  \BibitemOpen
  \bibfield  {author} {\bibinfo {author} {\bibfnamefont {M.}~\bibnamefont
  {D\c{a}browski}}\ and\ \bibinfo {author} {\bibfnamefont {K.}~\bibnamefont
  {Marosek}},\ }\href {\doibase 10.1007/s10714-018-2482-1} {\bibfield
  {journal} {\bibinfo  {journal} {General Relativity and Gravitation}\ }\textbf
  {\bibinfo {volume} {50}},\ \bibinfo {pages} {160} (\bibinfo {year}
  {2018})}\BibitemShut {NoStop}%
\bibitem [{\citenamefont {Barrow}(2004{\natexlab{a}})}]{BarrowSFS}%
  \BibitemOpen
  \bibfield  {author} {\bibinfo {author} {\bibfnamefont {J.~D.}\ \bibnamefont
  {Barrow}},\ }\href {\doibase 10.1088/0264-9381/21/11/l03} {\bibfield
  {journal} {\bibinfo  {journal} {Classical and Quantum Gravity}\ }\textbf
  {\bibinfo {volume} {21}},\ \bibinfo {pages} {L79} (\bibinfo {year}
  {2004}{\natexlab{a}})}\BibitemShut {NoStop}%
\bibitem [{\citenamefont {Barrow}(2004{\natexlab{b}})}]{GSFS-Barrow}%
  \BibitemOpen
  \bibfield  {author} {\bibinfo {author} {\bibfnamefont {J.~D.}\ \bibnamefont
  {Barrow}},\ }\href {\doibase 10.1088/0264-9381/21/23/020} {\bibfield
  {journal} {\bibinfo  {journal} {Classical and Quantum Gravity}\ }\textbf
  {\bibinfo {volume} {21}},\ \bibinfo {pages} {5619} (\bibinfo {year}
  {2004}{\natexlab{b}})}\BibitemShut {NoStop}%
\bibitem [{\citenamefont {Caldwell}\ and\ \citenamefont
  {Kamionkowski}(2004)}]{Caldwell_2004}%
  \BibitemOpen
  \bibfield  {author} {\bibinfo {author} {\bibfnamefont {R.~R.}\ \bibnamefont
  {Caldwell}}\ and\ \bibinfo {author} {\bibfnamefont {M.}~\bibnamefont
  {Kamionkowski}},\ }\href {\doibase 10.1088/1475-7516/2004/09/009} {\bibfield
  {journal} {\bibinfo  {journal} {Journal of Cosmology and Astroparticle
  Physics}\ }\textbf {\bibinfo {volume} {2004}},\ \bibinfo {pages} {009}
  (\bibinfo {year} {2004})}\BibitemShut {NoStop}%
\bibitem [{\citenamefont {Dunajski}\ and\ \citenamefont
  {Gibbons}(2008)}]{Dunajski_2008}%
  \BibitemOpen
  \bibfield  {author} {\bibinfo {author} {\bibfnamefont {M.}~\bibnamefont
  {Dunajski}}\ and\ \bibinfo {author} {\bibfnamefont {G.}~\bibnamefont
  {Gibbons}},\ }\href {\doibase 10.1088/0264-9381/25/23/235012} {\bibfield
  {journal} {\bibinfo  {journal} {Classical and Quantum Gravity}\ }\textbf
  {\bibinfo {volume} {25}},\ \bibinfo {pages} {235012} (\bibinfo {year}
  {2008})}\BibitemShut {NoStop}%
\bibitem [{\citenamefont {D\c{a}browski}(2005)}]{MPD2005}%
  \BibitemOpen
  \bibfield  {author} {\bibinfo {author} {\bibfnamefont {M.~P.}\ \bibnamefont
  {D\c{a}browski}},\ }\href {\doibase
  https://doi.org/10.1016/j.physletb.2005.08.080} {\bibfield  {journal}
  {\bibinfo  {journal} {Physics Letters B}\ }\textbf {\bibinfo {volume}
  {625}},\ \bibinfo {pages} {184} (\bibinfo {year} {2005})}\BibitemShut
  {NoStop}%
\bibitem [{\citenamefont {Bassett}\ \emph {et~al.}(2006)\citenamefont
  {Bassett}, \citenamefont {Tsujikawa},\ and\ \citenamefont {Wands}}]{Bassett}%
  \BibitemOpen
  \bibfield  {author} {\bibinfo {author} {\bibfnamefont {B.~A.}\ \bibnamefont
  {Bassett}}, \bibinfo {author} {\bibfnamefont {S.}~\bibnamefont {Tsujikawa}},
  \ and\ \bibinfo {author} {\bibfnamefont {D.}~\bibnamefont {Wands}},\ }\href
  {\doibase 10.1103/RevModPhys.78.537} {\bibfield  {journal} {\bibinfo
  {journal} {Rev. Mod. Phys.}\ }\textbf {\bibinfo {volume} {78}},\ \bibinfo
  {pages} {537} (\bibinfo {year} {2006})}\BibitemShut {NoStop}%
\bibitem [{\citenamefont {Mukhanov}(2005)}]{Mukhanov}%
  \BibitemOpen
  \bibfield  {author} {\bibinfo {author} {\bibfnamefont {V.}~\bibnamefont
  {Mukhanov}},\ }\href@noop {} {\emph {\bibinfo {title} {{Physical Foundations
  of Cosmology}}}}\ (\bibinfo  {publisher} {Cambridge Univ. Press},\ \bibinfo
  {address} {Cambridge},\ \bibinfo {year} {2005})\BibitemShut {NoStop}%
\bibitem [{\citenamefont {Olver}\ \emph {et~al.}(2010)\citenamefont {Olver},
  \citenamefont {Lozier}, \citenamefont {Boisvert},\ and\ \citenamefont
  {Clark}}]{NIST}%
  \BibitemOpen
  \bibfield  {author} {\bibinfo {author} {\bibfnamefont {F.~W.}\ \bibnamefont
  {Olver}}, \bibinfo {author} {\bibfnamefont {D.~W.}\ \bibnamefont {Lozier}},
  \bibinfo {author} {\bibfnamefont {R.~F.}\ \bibnamefont {Boisvert}}, \ and\
  \bibinfo {author} {\bibfnamefont {C.~W.}\ \bibnamefont {Clark}},\ }\href@noop
  {} {\emph {\bibinfo {title} {NIST Handbook of Mathematical Functions}}},\
  \bibinfo {edition} {1st}\ ed.\ (\bibinfo  {publisher} {Cambridge University
  Press},\ \bibinfo {address} {USA},\ \bibinfo {year} {2010})\BibitemShut
  {NoStop}%
\bibitem [{\citenamefont {Aghanim}\ \emph
  {et~al.}(2020{\natexlab{b}})\citenamefont {Aghanim}, \citenamefont {Akrami},
  \citenamefont {Ashdown}, \citenamefont {Aumont}, \citenamefont {Baccigalupi},
  \citenamefont {Ballardini}, \citenamefont {Banday}, \citenamefont {Barreiro},
  \citenamefont {Bartolo},\ and\ \citenamefont {et~al.}}]{Planck}%
  \BibitemOpen
  \bibfield  {author} {\bibinfo {author} {\bibfnamefont {N.}~\bibnamefont
  {Aghanim}}, \bibinfo {author} {\bibfnamefont {Y.}~\bibnamefont {Akrami}},
  \bibinfo {author} {\bibfnamefont {M.}~\bibnamefont {Ashdown}}, \bibinfo
  {author} {\bibfnamefont {J.}~\bibnamefont {Aumont}}, \bibinfo {author}
  {\bibfnamefont {C.}~\bibnamefont {Baccigalupi}}, \bibinfo {author}
  {\bibfnamefont {M.}~\bibnamefont {Ballardini}}, \bibinfo {author}
  {\bibfnamefont {A.~J.}\ \bibnamefont {Banday}}, \bibinfo {author}
  {\bibfnamefont {R.~B.}\ \bibnamefont {Barreiro}}, \bibinfo {author}
  {\bibfnamefont {N.}~\bibnamefont {Bartolo}}, \ and\ \bibinfo {author}
  {\bibnamefont {et~al.}},\ }\href {\doibase 10.1051/0004-6361/201833910}
  {\bibfield  {journal} {\bibinfo  {journal} {Astronomy \& Astrophysics}\
  }\textbf {\bibinfo {volume} {641}},\ \bibinfo {pages} {A6} (\bibinfo {year}
  {2020}{\natexlab{b}})}\BibitemShut {NoStop}%
\bibitem [{\citenamefont {Fixsen}(2009)}]{Fixsen2009}%
  \BibitemOpen
  \bibfield  {author} {\bibinfo {author} {\bibfnamefont {D.~J.}\ \bibnamefont
  {Fixsen}},\ }\href {\doibase 10.1088/0004-637x/707/2/916} {\bibfield
  {journal} {\bibinfo  {journal} {The Astrophysical Journal}\ }\textbf
  {\bibinfo {volume} {707}},\ \bibinfo {pages} {916} (\bibinfo {year}
  {2009})}\BibitemShut {NoStop}%
\end{thebibliography}%

\end{document}